\documentclass[%
 aip,
 jmp,
 amsmath,amssymb,
 reprint,%
author-year,
]{revtex4-1}

\usepackage{hyperref}% add hypertext capabilities

\usepackage{accents}
\newcommand{\utilde}[1]{\underaccent{\tilde}{#1}}
\usepackage [english]{babel}

\usepackage{tikz-cd}
\usetikzlibrary{babel}

\usepackage{dcolumn}% Align table columns on decimal point
\usepackage{bm}% bold math
\usepackage[utf8]{inputenc}
\usepackage[T1]{fontenc}
\usepackage{mathptmx}
\usepackage{etoolbox}

\usepackage [autostyle, english = american]{csquotes}
\MakeOuterQuote{"}

\makeatletter
\def\@email#1#2{%
 \endgroup
 \patchcmd{\titleblock@produce}
  {\frontmatter@RRAPformat}
  {\frontmatter@RRAPformat{\produce@RRAP{*#1\href{mailto:#2}{#2}}}\frontmatter@RRAPformat}
  {}{}
}%
\makeatother
\begin{document}

\preprint{AIP/123-QED}

\title[Relativistic massive particle with spin-1/2, a vector bundle point of view]{Relativistic massive particle with spin-1/2, a vector bundle point of view}
% Force line breaks with \\
\author{Heon Lee}
 \affiliation{Department of Mathematical Sciences, Seoul National University, 1, Gwanak-ro, Gwanak-gu, 08826 Seoul, Republic of Korea}%Lines break automatically or can be forced with \\
 \email{heoney93@gmail.com}

\date{Submitted July 22, 2021. Revised Nov 15, 2021}% It is always \today, today,

\begin{abstract}
Recently, in the context of Relativistic Quantum Information Theory (RQI) of massive spin-1/2 particles, it has been suggested that it is impossible to perform a momentum-independent spin measurement, showing the inadequacy of the spin reduced density matrix as a legitimate information resource. This is because there is an unavoidable ambiguity in the definition of the spin of a moving particle. In this paper, by introducing a vector bundle theoretic way to view the single-particle state space, we try to rule out this ambiguity. The discrete degree of freedom of the resulting representation space contains information about the Pauli-Lubansky four-vector of the particle instead of the ambiguous spin. Comparing this representation with the standard one used in the RQI literature, we show that the discrete degree of freedom of the standard representation space attains the meaning of the Newton-Wigner spin. Also using this viewpoint, we give a mathematical proof of why the spin reduced density matrix is meaningless, which is stronger than the previous claims in that it asserts that the matrix is void of any meaning at all, not just in terms of the impossibility of measurement or Lorentz non-covariance. We give a way (which turns out to be the only way) to modify it to obtain the Pauli-Lubansky reduced density matrix, which is covariant under Lorentz transformations. 
\end{abstract}

\maketitle

\section{\label{sec:1}Introduction}
The spin of a massive spin-1/2 particle is a primary resource that can be used as a qubit carrier. In a scenario where it is exploited in this way, one usually discards the momentum information of the particle and takes only the spin information of the particle. However, in the context of Special Relativity, this practice poses a serious problem.

In the seminal paper \cite{peres2002}, the authors showed that the reduced spin density matrix, which is obtained by taking the partial trace with respect to the momentum variable, does not transform covariantly under Lorentz transformations. Therefore, the notion "spin state of a particle independent of momentum" is relativistically meaningless. This work has generated intense studies (see \cite{peres2004, alsing2012} and references therein).

This phenomenon is due to the momentum dependence of the Wigner rotation in the transformation law Eq.~(\ref{eq:2}). But, more fundamentally, this stems from the inherent momentum dependence of the spin variable of the standard representation space, of which the Wigner rotation is just a manifestation. This fact is most clearly seen when we look more closely at the derivation of the standard irreducible representation space of the universal covering group of the Poincar\'e group, which has been predominantly used in the RQI literature to describe a massive spin-1/2 particle.

The standard representation space is a tensor product system
\begin{equation}\label{eq:1}
\mathcal{H} := L^2 \left(\mathbb{R}^3 , \mu \right) \otimes \mathbb{C}^2
\end{equation}
where $d \mu (\mathbf{p}) = \frac{d^3 \mathbf{p}}{ (2 \pi )^3 \sqrt{ \mathbf{p}^2 + m^2 }}$ is a Lorentz invariant measure on the mass shell and $L^2 \left(\mathbb{R}^3 , \mu \right)$ is the space of momentum wave functions, on which a Lorentz transformation $(\Lambda, a) \in \mathbb{R}^4 \ltimes SL(2, \mathbb{C})$ acts as
\begin{equation}\label{eq:2}
\left[U(\Lambda, a) \psi \right] (p) = e^{-i p_\mu a^\mu} W(\Lambda , \Lambda^{-1} p)  \psi (\Lambda^{-1} p ),
\end{equation}
where $\mathbf{p}\in \mathbb{R}^3$ is identified with $ p = ( \sqrt{ | \mathbf{p}| ^2 + m^2 } , \mathbf{p}) \in \mathbb{R}^4$, $W(\Lambda , p) = L(\Lambda p)^{-1} \Lambda L(p) \in SU(2)$ is the Wigner rotation matrix, and $L(p) \in SL(2, \mathbb{C})$ a boosting which maps $k = (m, 0, 0 , 0)$ to $p$.\footnote{These standard descriptions are from \cite{weinberg}.}

Usually (cf. \cite{weinberg}), one arrives at the conclusion Eqs.~(\ref{eq:1}), (\ref{eq:2}) by postulating

\begin{equation}\label{eq:3}
| p , \pm \rangle = U(L(p)) | k , \pm \rangle
\end{equation}
where $\pm$ denotes the spin-up/down states respectively. In effect, this assumption forces that the boosting $L(p)$ does not change the spin-up/down states. This assumption is quite arbitrary since $L(p)$ may very well be chosen as $L(p)B(p)$ for any rotation $B(p) \in SU(2)$ since $SU(2)$ fixes $k$, in which case the meaning of $\pm$ becomes obscure.

It seems that \cite{avelar2013} is the first paper where this arbitrariness is recognized. In the paper, the author exploited this as a freedom to generate a variety of reduced spin density matrices from which one can choose, to each physical setting, what is expected to give the most useful spin information. \cite{he2007, aghaee2017} used this freedom to obtain the reduced density matrix for helicity (the spin component along the direction of momentum) by choosing $L(p) R( \hat{\mathbf{p}})$ in place of $L(p)$ (here, $R(\hat{\mathbf{p}}) \in SU(2)$ is a rotation that maps $\hat{z}$ to $\hat{\mathbf{p}}$).

In fact, even the standard choice of $L (p)$ (cf. Eq.~(\ref{eq:19})) is troublesome since it would change the quantization axis of the spin by
\begin{equation}\label{eq:4}
(0, \mathbf{n}) \mapsto L(p) (0, \mathbf{n})
\end{equation}
which is not equal to $\mathbf{n}$ in general (cf. \cite{derbarba2012}). Hence the $\pm$ in the momentum-spin eigenstate Eq.~(\ref{eq:3}) in fact means that the state is in the definite spin-up/down states with respect to the changed axis $L(p) \hat{z}$ and not with respect to the $\hat{z}$-axis of a fixed inertial frame. Because of this fact, when we have no access to the momentum variable, the mere information of $\mathbb{C}^2$ component in Eq.~(\ref{eq:1}) does not give us any information about the spin at all, let alone the reduced spin density matrix which is formed by summing over these pieces of information \cite{saldanha2012a}.

The point is that, since there is no such thing as \emph{the} particle rest frame, one must make a certain choice of $L(p)$ to speak of the spin state of the particle, because of which this ambiguity in the meaning of spin becomes unavoidable.

This raises a question. Is there a way to rule out the stated arbitrariness of $L(p)$ so that the qubit information stored in the $\mathbb{C}^2$ component (e.g. $\pm$ in $|p, \pm \rangle$) would have meaning without recourse to the change of reference frames by $L(p)$? More precisely, we would like to have a kind of quantum coordinate system $(p, \chi) \in \mathbb{R}^4 \times \mathbb{C}^2$ for a fixed inertial observer which encodes the 4-momentum $p$ as well as the internal quantum state $\chi$ of a single particle such that $\chi \in \mathbb{C}^2$ itself would correctly reflect the perception of the fixed observer.

This question may be addressed only through the vector bundle description of single-particle states since removing the stated arbitrariness of $L(p)$ would break the tensor product structure of the representation space Eq.~(\ref{eq:1}) whereas, in the formalism of the bundle theory, the $\mathbb{C}^2$ component remains intact even after this removal. The key idea is to view the wave functions in Eq.~(\ref{eq:1}) as sections of a trivial vector bundle with fiber $\mathbb{C}^2$ over the mass shell Eq.~(\ref{eq:7}).

A vector bundle over a topological space $M$ can be thought of as a disjoint union $E:= \cup_{p\in X} E_p$ of vector spaces $E_p$ for each $p \in M$ endowed with a topology which assembles the vector spaces in a coherent manner.\footnote{Those who are not familiar with the vector bundle theory are referred to \cite{lee}, Chapter 10. But, only minimum knowledge of the theory will be needed in this paper.} A section of the bundle is a function $f:M \rightarrow E$ such that $f (p) \in E_p$ for each $p \in M$. In terms of this language, a wave function describing a massive particle with spin-1/2 is a section of the bundle $E=X \times \mathbb{C}^2$ over the mass shell $X$. So, we have a two-level quantum system $E_p$ for each momentum $p \in X$ and a wave function becomes a field of qubits that is a function of $p$.

As we shall see, rephrasing the usual wave function descriptions of Relativistic Quantum Mechanics in the vector bundle language has several advantages. First, the mentioned arbitrariness of $L(p)$ is easily removed using this viewpoint (Sect.~\ref{sec:3C}) and it also reveals the precise meaning of the spin variable in relation to changes of inertial frames (Sect.~\ref{sec:3F}). Also, it helps us to see more clearly why the spin reduced density matrix introduced in \cite{peres2002} has no real meaning as claimed in \cite{saldanha2012a} (Sect.~\ref{sec:4B}) and gives a way to modify it to obtain a Lorentz covariant reduced matrix (Sect.~\ref{sec:4C}).

The present paper is organized as follows. Sect.~\ref{sec:2} presents some notations and relations which are used throughout the paper. We introduce the promised vector bundle point of view in Sect.~\ref{sec:3}, where we deal with two kinds of bundles, one of which depends on the choice of $L(p)$ and the other completely free from it. Then, we investigate what information one can extract from the qubits in the respective fibers $\mathbb{C}^2$ of the two bundles.

There is a nonstandard representation space that naturally comes out from the latter bundle, which is equivalent to the standard one (Eqs.~(\ref{eq:1}) and (\ref{eq:2})), yet is free from the choice of $L(p)$. Section \ref{sec:4} is about this alternative representation space. Some features of this space are presented. In particular, we analyze the original spin reduced density matrix of \cite{peres2002} from the point of view introduced in the preceding section and show why it is void of any meaning at all, not just in terms of Lorentz non-covariance as noticed in \cite{peres2002} and the impossibility of measurement as noticed in \cite{saldanha2012a}, respectively. We also show that the construction of the Lorentz covariant reduced spin density matrix introduced in \cite{caban2005} is the only way to make the original one meaningful.

I conclude the paper with some remarks (Sect.~\ref{sec:5}) on the alternative representation space, where some previous treatments of it in the RQI literature can be found.

\section{\label{sec:2}Preliminaries}

In this section, let's set some notations and mathematical facts which will be used throughout. First, we use the Minkowski metric
\begin{equation}\label{eq:5}
\eta = \text{diag} (1, -1, -1,-1)
\end{equation}
and the Pauli matrices
\begin{equation}\label{eq:6}
\tau^0 = I, \hspace{0.1cm} \tau^1 = \begin{pmatrix} 0 & 1 \\ 1 & 0 \end{pmatrix}, \hspace{0.1cm} \tau^2 = \begin{pmatrix} 0 & -i \\ i & 0 \end{pmatrix}, \hspace{0.1cm}  \tau^3 = \begin{pmatrix} 1 & 0 \\ 0 & -1 \end{pmatrix}.
\end{equation}

Fix a positive number $m>0$. Consider the following diffeomorphism from $\mathbb{R}^3$ onto the \textit{mass shell}
\begin{equation}\label{eq:7}
X := \left\{ p \in \mathbb{R}^4 : p^0 > 0 , \hspace{0.1cm} (p)^2 = m^2 \right\}
\end{equation}
given by
\begin{equation}\label{eq:8}
\mathbf{p} \mapsto \left(\sqrt{m^2 + \mathbf{p}^2 } , \mathbf{p} \right).
\end{equation}

Denoting $p^0 = \sqrt{ m^2 + \mathbf{p}^2 }$, we will always identify $\mathbf{p}$ with the four-vector $p^\mu = (p^0 , \mathbf{p}) \in X$ via this diffeomorphism throughout this paper even though it is not explicitly stated (in fact, this was already done in Eq.~(\ref{eq:2})). Also, we fix $k= (m,0,0,0) \in X$.

$X$ is a Lorentz invariant subset, i.e., $\Lambda p \in X$ for all $p \in X$ and $\Lambda \in SO^\uparrow (1,3)$ ($SO^\uparrow (1,3)$ is the connected component of $O(1,3)$). The measure $\mu$ on $X$ defined by
\begin{equation}\label{eq:9}
d \mu (p) = \frac{d^3 \mathbf{p}}{ (2 \pi)^3 \sqrt{ m^2 + \mathbf{p}^2 }}
\end{equation}
is a Lorentz invariant measure (cf. \cite{folland2008}), that is, for all integrable functions $f$ on $X$ and $\Lambda \in SO^\uparrow (1,3)$,
\begin{equation}\label{eq:10}
\int_X f (\Lambda p ) d \mu (p) = \int_X f (p) d \mu (p).
\end{equation}

Note that each $p \in X$ satisfies the following relation.
\begin{equation}\label{eq:11}
p_\mu p^\mu = m^2
\end{equation}

Denote, for $x \in \mathbb{R}^4$,
\begin{subequations}\label{eq:12}
\begin{equation}\label{eq:12a}
\utilde{x} = \begin{pmatrix} x_0 + x_3 & x_1 - i x_2 \\ x_1 + i x_2 & x_0 - x_3  \end{pmatrix} = x_{\mu} \tau^{\mu}
\end{equation}
\begin{equation}\label{eq:12b}
\tilde{x} = \begin{pmatrix} x^0 + x^3 & x^1 - i x^2 \\ x^1 + i x^2 & x^0 + x^3 \end{pmatrix} = x^0 \tau^0 + \mathbf{x} \cdot \boldsymbol{\tau}
\end{equation}
\end{subequations}
where of course the usual convention of index raising and lowering via $\eta$ as well as Einstein's summation convention is in action (These notations are borrowed from the book \cite{bleecker} with a slight modification).

A direct calculation shows
\begin{equation}\label{eq:13}
\utilde{x}\tilde{y} + \utilde{y} \tilde{x} = 2 \langle x, y \rangle I = \tilde{x} \utilde{y} + \tilde{y} \utilde{x}
\end{equation}
where $\langle x, y \rangle = x_\mu y^\mu$. For $p \in X$, the two matrices $\tilde{p}$ and $\utilde{p}$ are positive matrices with the square roots given by
\begin{subequations}
\label{eq:14}
\begin{eqnarray}
\sqrt{\utilde{p} } = \frac{1}{\sqrt{ 2 (m + p_0 ) }} \left( \utilde{p} + m I \right) \\ 
\sqrt{\tilde{p} } = \frac{1}{\sqrt{ 2 (m + p_0 ) }} \left( \tilde{p} + m I \right)
\end{eqnarray}
\end{subequations}
meaning that these two matrices are positive and their squares are $\utilde{p}$ and $\tilde{p}$, respectively. It is easy to see
\begin{equation}\label{eq:15}
\sqrt{\utilde{p}} \sqrt{\tilde{p}} = mI = \sqrt{\tilde{p}}\sqrt{\utilde{p}},
\end{equation}
which may be expressed as
\begin{equation}\label{eq:16}
\left( \sqrt{\frac{\utilde{p}}{m}} \right)^{-1} = \left( \sqrt{\frac{\tilde{p}}{m}} \right).
\end{equation}

Let
\begin{equation}\label{eq:17}
\kappa : SL ( 2, \mathbb{C} ) \rightarrow SO^{\uparrow} (1,3)
\end{equation}
be the standard double covering homomorphism (cf. \cite{folland2008}) which is defined as, for an arbitrary four-vector $x^\mu \in \mathbb{R}^4$,
\begin{subequations}\label{eq:18}
\begin{eqnarray}
\left(\kappa (\Lambda) x \right)^\sim &=& \Lambda \tilde{x} \Lambda^*  \\
\left( \kappa (\Lambda) x \right)_\sim &=& \Lambda^{\dagger -1} \utilde{x} \Lambda^{-1} .
\end{eqnarray}
\end{subequations}

Given $\Lambda \in SL(2, \mathbb{C})$ and $ x \in \mathbb{R}^4$, we write $\kappa (\Lambda) x = \Lambda x$ and omit $\kappa$ from now on.

\subsection{\label{sec:2A}The standard boostings}

The standard choice for $L(p)$ has been (cf. \cite{weinberg})
\begin{equation}\label{eq:19}
L_0 (p) := R (\hat{\mathbf{p}} ) B( | \mathbf{p} |) R( \hat{\mathbf{p}} ) ^{-1}
\end{equation}
where, if $\hat{p} = ( \sin \theta \cos \phi, \sin \theta \sin \phi , \cos \theta )$, then
\begin{equation}\label{eq:20}
R(\hat{\mathbf{p}}) := \begin{pmatrix} e^{- i \frac{\phi}{2}} & 0 \\ 0 & e^{i \frac{\phi}{2}} \end{pmatrix} \begin{pmatrix} \cos \frac{\theta}{2} & - \sin \frac{\theta}{2} \\ \sin \frac{\theta}{2} & \cos \frac{\theta}{2} \end{pmatrix} \in SU(2)
\end{equation}
is a rotation which takes $\hat{z}$ to $\hat{\mathbf{p}}$ and
\begin{equation}\label{eq:21}
B( |\mathbf{p}|) := \begin{pmatrix} \sqrt{  \frac{p^0 + | \mathbf{p}|}{m} }    & 0 \\ 0 & \sqrt{  \frac{p^0 - | \mathbf{p}|}{m} } \end{pmatrix} \in SL(2, \mathbb{C})
\end{equation}
is the boosting along the $z$-axis which takes $k$ to $(p^0 , 0, 0 , | \mathbf{p}|)$.

Given the fact that $L_0 (p)$ is a positive matrix (since $R^{-1} = R^\dagger$ for $R \in SU(2)$), the calculation
\begin{align}\label{eq:22}
L_0 (p)^2 &= R(\hat{\mathbf{p}}) B( | \mathbf{p}|)^2 R(\hat{\mathbf{p}})^{-1} \nonumber \\
=& \frac{1}{m} R(\hat{\mathbf{p}})  \begin{pmatrix} p^0 + |\mathbf{p}| & 0 \\ 0 & p^0 - |\mathbf{p}| \end{pmatrix} R(\hat{\mathbf{p}})^\dagger 
= \frac{1}{m} \tilde{p}
\end{align}
shows, by the uniqueness of the matrix square root, that
\begin{equation}\label{eq:23}
L_0 (p) = \sqrt{ \frac{\tilde{p}}{m}}.
\end{equation}

\section{\label{sec:3} The vector bundle descriptions}
\subsection{\label{sec:3A}The trivial vector bundle}

Another way to view the representation space Eq.~(\ref{eq:1}) is to write it as the $L^2$-section space of the trivial bundle $E:=X \times \mathbb{C}^2 \xrightarrow{\xi} X$ over the mass shell $X$. Indeed, with the identification $\mathbb{R}^3 \cong X$ given in Eq.~(\ref{eq:8}) and the trivial Hermitian metric $g : E \otimes E \rightarrow \mathbb{C}$ defined by
\begin{equation}\label{eq:24}
g\left( (p,v) , (p, w) \right) = v^\dagger w,
\end{equation}
we see Eq.~(\ref{eq:1}) may be rewritten as
\begin{align}\label{eq:25}
\mathcal{H} = \Big\{ \psi : X \rightarrow E : \xi \circ \psi (p) = p \hspace{0.05cm} 
\& \int_X g( \psi , \psi) \mu < \infty \Big\}.
\end{align}

We write the fibers as $E_p := \{ p \} \times \mathbb{C}^2$. So, we have a two-level quantum system $E_p$ for each momentum $p \in X$. In our formalism, the momentum-spin eigenstate $|p, \pm \rangle$ can be identified with the point $ (p, |\pm \rangle ) \in E_p$ where $| + \rangle = \begin{pmatrix} 1 \\ 0 \end{pmatrix}$ and $|- \rangle = \begin{pmatrix} 0 \\ 1 \end{pmatrix}$.

\subsection{\label{sec:3B}The transformation law \textrm{I}}

Every inertial observer is given the bundle $E$ for the description of a massive spin-1/2 particle. How does the vector bundle description change under Lorentz transformations? Suppose an inertial observer (call her Alice) has prepared a single-particle in the state $\psi \in \mathcal{H}$. Consider another inertial observer (call him Bob) whose frame is Lorentz transformed by $(\Lambda , a) \in \mathbb{R}^4 \ltimes SL(2, \mathbb{C})$ from Alice's (which means that Alice's coordinate $x_A ^\mu$ and Bob's $x_B ^\mu$ are related by $x_B = a +  \Lambda x_A $). Then, according to the postulate of Relativistic Quantum Mechanics (cf. \cite{weinberg}), the particle is seen to be in the state $U(\Lambda,a) \psi \in \mathcal{H}$ in Bob's reference frame (cf. Eq.~(\ref{eq:2})).

For this to be true, Bob's vector bundle description and that of Alice should be related by the vector bundle isomorphism

\begin{gather}
V(\Lambda, a ) : E^A \rightarrow E^B \nonumber \\
 \quad (p,v) \mapsto \left(  \Lambda p, e^{- i (\Lambda p)_\mu a^\mu}W( \Lambda, p) v \right) \label{eq:26}
\end{gather}
so that
\begin{equation}\label{eq:27}
U(\Lambda, a ) \psi = V( \Lambda, a) \circ \psi \circ \Lambda^{-1}
\end{equation}
becomes the transformation law Eq.~(\ref{eq:2}) for the sections. The following commutative diagram will be useful in visualizing the transformation law
\begin{equation}\label{eq:28}
\begin{tikzcd}[baseline=(current  bounding  box.center)]
E^A \arrow[r, "{V( \Lambda, a)}"] \arrow[d, "\xi" ]
& E^B \arrow[d, "\xi"]  \\
X  \arrow[r, "\Lambda"]
& X 
\end{tikzcd}.
\end{equation}

Note that since $W(\Lambda ,p ) \in SU(2)$, $V(\Lambda, a)$ is an Hermitian bundle isomorphism.

\subsection{\label{sec:3C}An alternative vector bundle}

As can be seen from the presence of the Wigner rotation in the transformation law Eq.~(\ref{eq:26}), the $L(p)$-dependence of the $\mathbb{C}^2$ component is also present in this vector bundle. We set to eliminate it. We define an Hermitian metric $h$ on $X \times \mathbb{C}^2$ by

\begin{equation}\label{eq:29}
h\left( (p,v) , (p, w) \right) = v^\dagger \frac{ \utilde{p}}{m} w.
\end{equation}

We denote the Hermitian bundle $(X \times \mathbb{C}^2, h)$ as $E'$ and the trivial bundle $(X \times \mathbb{C}^2,g)$ of Sect.~\ref{sec:3A} as $E$. We distinguish elements of the two bundles by writing $'$ for the elements of the former bundle, e.g., $(p,v) \in E$ and $(p,v)' \in E'$ even though they are elements of the same underlying set. Then, the vector bundle isomorphism from $E$ onto $E'$ given by

\begin{equation}\label{eq:30}
(p,v) \mapsto (p, L(p)v)'
\end{equation}
is an Hermitian bundle isomorphism since every boosting $L(p)$ can be written as $L(p) = L_0 (p) B(p)$ for some rotation $B(p) \in SU(2)$ (This is because $X \cong SL(2, \mathbb{C})/SU(2)$, the left $SU(2)$-coset space) and hence
\begin{align}\label{eq:31}
L(p)^\dagger \frac{\utilde{p}}{m} L(p) =B(p)^\dagger L_0 (p)^\dagger L_0 (p)^{-2} L_s (p) B(p) \nonumber \\
= B(p)^\dagger B(p) = I_2
\end{align}
where we have used Eq.~(\ref{eq:23}). This Hermitian bundle isomorphism will be denoted as $L: E \rightarrow E'$.

\subsection{\label{sec:3D}The transformation law \textrm{II}}

Consider the situation at the beginning of Sect.~\ref{sec:3B}. This time, every inertial observer is given the bundle $E'$ for the description of a massive spin-1/2 particle. Using the isomorphism Eq.~(\ref{eq:30}), we see that the transformation law for the bundle $E'$ is given by

\begin{gather}
V'(\Lambda, a) := L \circ V(\Lambda, a ) \circ L^{-1} : E'^A \rightarrow E'^B \nonumber \\
(p,v)' \mapsto \left(\Lambda p , e^{-i (\Lambda p)_\mu a^\mu} \Lambda v \right)' \label{eq:32}
\end{gather}
\begin{equation}\label{eq:33}
\begin{tikzcd}[baseline=(current  bounding  box.center)]
E'^A \arrow[r, "{V'( \Lambda, a)}"] \arrow[d, "\xi" ]
& E'^B \arrow[d, "\xi"]  \\
X  \arrow[r, "\Lambda"]
& X 
\end{tikzcd}
\end{equation}
which is an Hermitian bundle isomorphism being a composition of three Hermitian bundle isomorphisms.

We see that the Wigner rotation has been removed from the transformation law and hence the bundle description $E'$ is free from the choice of $L(p)$.

\subsection{\label{sec:3E}The Pauli-Lubansky four-vector and the Newton-Wigner spin}

To give the vector bundle description a physical interpretation, we need to digress a little bit to discuss the \textit{Pauli-Lubansky four-vector}. The qubits in the bundle $E'$ will be shown to contain information about the Pauli-Lubansky four-vector of the particle, whereas those in the bundle $E$, once the choice $L(p)=L_0 (p)$ is made, will have information about the \textit{Newton-Wigner spin} (cf. Sections~\ref{sec:3F} and \ref{sec:4A}).

Suppose a particle has a relativistic momentum and angular momentum $p^\mu$ and $j_{\alpha \beta}$, respectively (cf. \cite{anderson}). Then, the Pauli-Lubansky four-vector of the particle is defined as

\begin{equation}\label{eq:34}
w^\mu = \frac{1}{2} \varepsilon^{\nu \alpha \beta \mu} p_{\nu} j_{\alpha \beta}
\end{equation}
which is related to spin as follows. First, the following properties are easily seen from the definition.
\begin{equation}\label{eq:35}
p_\mu w^\mu = 0
\end{equation}
and when $p= (m,0,0,0)$,
\begin{equation}\label{eq:36}
w = (0, m \boldsymbol{j}).
\end{equation}

Using Eqs.~(\ref{eq:12}), (\ref{eq:13}), (\ref{eq:18}), (\ref{eq:23}) and (\ref{eq:35}), we calculate
\begin{align}\label{eq:37}
\left(L_0 (p)^{-1} w \right)^\sim =  \sqrt{ \frac{\utilde{p}}{m}} \tilde{w} \sqrt{\frac{\utilde{p}}{m}}
 = \frac{1}{\sqrt{ 2m (m + p^0) }} ( \utilde{p} + mI ) \tilde{w} \sqrt{\frac{\utilde{p}}{m}}  \nonumber \\
= \frac{1}{\sqrt{ 2m (m + p^0 ) }} (2 \langle p,w \rangle - \utilde{w} \tilde{p} + 2 m w^0 I - m \utilde{w}) \sqrt{\frac{\utilde{p}}{m}} \nonumber \\
= \frac{1}{2m(m+p^0)}  ( 2m w^0 (\utilde{p} + mI) ) - \utilde{w} \nonumber  \\
=  \frac{1} {m+ p^0 } w^0 ( \utilde{p} + mI) - \utilde{w}
\end{align}
from which we see that $L_0 (p)^{-1} w = (0, m \mathbf{s})$ where
\begin{equation}\label{eq:38}
\boldsymbol{s} = \frac{1}{m} \left( \mathbf{w} - \frac{ w^0 \mathbf{p}}{m +p^0} \right)
\end{equation}
is called the \textit{Newton-Wigner spin three-vector}.

Thus, the Newton-Wigner spin for a particle with momentum $p$ and angular momentum $j$ is just the Pauli-Lubansky vector seen from an $L_0(p)^{-1}$-transformed inertial observer, with respect to whom the particle is at rest consequently. Therefore, together with Eq.~(\ref{eq:36}), it has the meaning of the internal angular momentum, i.e., the spin.\footnote{Eq.~(\ref{eq:38}) is also what appears in the relativistic treatment of the interaction between spin and electromagnetic field (cf. \cite{anderson}, pp.248--253).} Note that the definition of the Newton-Wigner spin depends on the choice of the boostings $L(p) = L_0 (p)$.

For a general boosting $L(p) = L_0 (p) B(p)$, we have
\begin{equation}\label{eq:39}
L(p)^{-1} w = B(p)^{-1} (0, m \mathbf{s}) = (0, m [B(p)^{-1} \mathbf{s} ]).
\end{equation}
and hence $w$ always satisfies
\begin{equation}\label{eq:40}
w = L(p) (0, m \mathbf{v})
\end{equation}
for some three-vector $\mathbf{v}$ such that $|\mathbf{v}| = |\mathbf{s}|$ regardless of the choice of $L(p)$.
 
So, the Pauli-Lubansky four-vector is just the spin vector as seen from an observer who perceives the spin-carrying particle as moving with momentum $p$, and this interpretation (as well as the definition Eq.~(\ref{eq:34})) is independent of the choice of $L(p)$ in contrast to the case of the Newton-Wigner spin.

This asymmetry of $L(p)$-dependence between the Pauli-Lubansky vector and the Newton-Wigner spin is the classical analogue of the asymmetry between the $E$-bundle and $E'$-bundle descriptions, as we shall see in the next section.

\subsection{\label{sec:3F}A physical interpretation}

In this section, the promised interpretations given at the beginning of the preceding section are presented.

\subsubsection{\label{sec:3F1}The \texorpdfstring{$E$}{TEXT}-bundle description}

In the Quantum Mechanics of the two-level system, if $\chi \in \mathbb{C}^2$ is a qubit, then it is the spin-up state along the direction $\mathbf{n}:= \chi^\dagger \boldsymbol{\tau} \chi$. That is, we have
\begin{equation}\label{eq:41}
(\boldsymbol{\tau}\cdot \mathbf{n}) \chi = \chi
\end{equation}
and in fact,
\begin{equation}\label{eq:42}
\boldsymbol{\tau}\cdot \mathbf{n} = \chi \chi^\dagger - (I_2 - \chi \chi^\dagger) = 2 \chi \chi^\dagger - I_2
\end{equation}
since a state orthogonal to $\chi$ is the spin-down state along the direction $\mathbf{n}$. So, we see that the state $\chi$ is completely characterized by the three-vector $\mathbf{n}$ and we may call it the \textit{spin direction} of $\chi$.

Since it will become very important in Sect.~\ref{sec:4}, we rewrite this result as
\begin{equation}\label{eq:43}
\chi \chi^\dagger = \frac{1}{2} ( \boldsymbol{\tau}\cdot \mathbf{n} +I_2 ).
\end{equation}

Fix an inertial observer Bob. Consider an $L(p)^{-1}$-transformed observer Alice who has prepared a qubit $(k,\chi) \in E_k ^A$ so that the vector $\mathbf{n}$ formed as in Eq.~(\ref{eq:42}) in her frame is what Alice perceives as the information content of the qubit.

Since
\begin{equation}\label{eq:44}
V(L(p)) (k,\chi) = (p, \chi),
\end{equation}
according to Eq.~(\ref{eq:26}), we see that the qubit is seen as $(p, \chi) \in E_p ^B$ in Bob's frame. As Alice perceives the qubit as the three-vector $\mathbf{n}$ in her frame, Bob must see this qubit as the four-vector $L(p) (0, \mathbf{n})$. However, if Bob is not provided with the knowledge of $L(p)$, then the only information that Bob could get from $(p, \chi) \in E_p ^B$ is the vector $\mathbf{n}$ which is not what he perceives as the spin direction in his frame.

So, we see that the qubits in $E_p ^B$ don't reflect Bob's perception of the spin state. Rather, they are the perception of the $L(p)^{-1}$-transformed observer Alice. This shows that the problem raised in the introduction (see the paragraphs containing Eqs.~(\ref{eq:3}) and (\ref{eq:4})) still resides in the $E$-bundle description. I.e., the vectors contained in $E_p ^B$ themselves don't have a meaning in Bob's reference frame. They become useful only if Bob is also provided with the knowledge of $L(p)$.

In this sense, the $E$-bundle description is similar to the nature of the Newton-Wigner spin where one had to choose $L(p) = L_0 (p)$ to define it. This point will be revisited in Sect.~\ref{sec:3F3} and also in Sect.~\ref{sec:4A}.

\subsubsection{\label{sec:3F2}The \texorpdfstring{$E'$}{TEXT}-bundle description}

We might ask, what about $(p, \chi)' \in E'^B _p $? Does $\chi$ itself have a meaning independent of the choice of $L(p)$ unlike the case of the $E$-bundle? We assert that this is so, by showing that we can extract information about the Pauli-Lubansky four-vector of the qubit without the knowledge of $L(p)$.

Let $(p, \chi) \in E'_p$ with $\| L(p)^{-1} \chi \|_{E_p} = \| \chi \|_{E'_p} = 1$ and hence it can be called a qubit. Because of this condition and Eqs.~(\ref{eq:43}) and (\ref{eq:18}), we have
\begin{equation}\label{eq:45}
m \chi \chi^\dagger - \frac{ \tilde{p}}{2} = m L(p) \left( \frac{1}{2}  \boldsymbol{\tau}\cdot \mathbf{n} \right) L(p) = \frac{m}{2} (L(p) (0, \mathbf{n}))^\sim
\end{equation}
where $\mathbf{n}$ is the spin direction of the qubit $L(p)^{-1} \chi \in \mathbb{C}^2$. So, there is a four-vector $w \in \mathbb{R}^4$ such that
\begin{equation}\label{eq:46}
\tilde{w} = m \chi \chi^\dagger - \frac{ \tilde{p}}{2}
\end{equation}
which will be called the \textit{Pauli-Lubansky four-vector of the qubit $(p, \chi)' \in E'_p$}. Note that this definition is completely free from any reference to the boostings $L(p)$. $L(p)$ came into the picture for the sole purpose of showing that the RHS of Eq.~(\ref{eq:46}) can be represented by a four-vector as in Eq.~(\ref{eq:12b}).

Let's see if there is a link between this definition and the Pauli-Lubansky four-vector of the particle defined in Sect.~\ref{sec:3E}. As in Sect.~\ref{sec:3F1}, Bob is a fixed inertial observer and Alice is an $L(p)^{-1}$-transformed observer with respect to Bob. She prepares a qubit $(k, \chi)' \in E'^A_k$ in her rest frame. Observe
\begin{equation}\label{eq:47}
V'(L(p) ) (k, \chi)'= (p, L(p) \chi)'
\end{equation}
and hence from Bob's frame, the qubit looks like $(p, L(p) \chi)' \in E'_p$. He forms the Pauli-Lubansky vector for the qubit $(p, L(p)\chi)' \in E'_p$ according to Eq.~(\ref{eq:46}) to find that
\begin{equation}\label{eq:48}
\tilde{w} = ( L(p) ( 0, \frac{\mathbf{n}}{2}) )^\sim
\end{equation}
where $\mathbf{n}$ is the spin direction of the qubit $(k, \chi)' \in E'^A_k$ in Alice's frame.

So, the Pauli-Lubansky four-vector of the qubit, Eq.~(\ref{eq:46}), really has the meaning of the Pauli-Lubansky four-vector of the particle (cf. Eq.~(\ref{eq:40})) if we accept the usual interpretation of the qubits in the rest frame (i.e. those in $E'^A _k$) as giving information about the spin three-vector of the particle.

In this sense, one may call the qubits in the $E'$-bundle "apparent information" since they are directly accessible from the rest frame without recourse to frame changes, whereas those in the $E$-bundle are not, except those in the fiber $E_k$. In this regard, the bundle $E'$ is more suitable as a "quantum coordinate system for a single particle" than the bundle $E$ and this is the main reason behind the introduction of the vector bundle point of view as mentioned in the introduction.

\subsubsection{\label{sec:3F3}The \texorpdfstring{$E_0$}{TEXT}-bundle description}

Before we conclude this section, let's revisit the $E$-bundle description of Sect.~\ref{sec:3F1}. But, this time, we make a choice $L(p) = L_0 (p)$ throughout this subsection. The $E$-bundle in this case is denoted as $E_0$. This case is of special interest since the Hilbert space description constructed from $E_0$ has been the most thoroughly studied case in the RQI literature.

Taking the interpretation given in Sect.~\ref{sec:3F2} for granted, a qubit $(p, \chi)' \in E'_p$ has information about the Pauli-Lubansky four-vector of the particle. The corresponding vector in $(E_0)_p$ is $(p, L_0 (p)^{-1} \chi)$ (cf. Eq.~(\ref{eq:30})). Since $\| L_0 (p)^{-1} \chi \|_{E_p} = \| \chi \|_{E'_p} = 1$, it is a qubit in a usual sense. Form the spin three-vector $\mathbf{n}$ of this qubit as in Eq.~(\ref{eq:43}). Then, by Eq.~(\ref{eq:46}) and Eq.~(\ref{eq:37}),

\begin{align}\label{eq:49}
\boldsymbol{\tau} \cdot \left(\frac{1}{2}  \mathbf{n}\right) &=  L_0 (p)^{-1} \chi \chi^\dagger L_0 (p)^{-1} - \frac{1}{2} I_2 \nonumber \\
&=  L_0 (p)^{-1} \left( \frac{\tilde{w}}{m} + \frac{\tilde{p}}{2m}  \right) L_0 (p)^{-1} - \frac{1}{2} I_2 \nonumber \\
& = \frac{1}{m} \left(L_0 (p)^{-1} w \right)^\sim = ((0,\mathbf{s}))^\sim = \boldsymbol{\tau} \cdot \mathbf{s}
\end{align}
where $\mathbf{s}$ is the Newton-Wigner spin defined in Eq.~(\ref{eq:38}).

So, we conclude that the information contained in the qubits of the bundle $E _0$ in relation to those in $E'$ is precisely the Newton-Wigner spin of the particle. This interpretation will be strengthened in Sect.~\ref{sec:4A} where we show that a similar account also holds on the level of wave functions and operators.

\section{\label{sec:4}An alternative representation space}

Let $\mathcal{H}'$ be the $L^2$-section space of the Hermitian bundle $E'$. I.e., $\mathcal{H}'$ is the set of all measurable sections $f : X \rightarrow E'$ which satisfy

\begin{equation}\label{eq:50}
\int_X   h\left(f, f\right) \mu < \infty .
\end{equation}

Then, $\mathcal{H}'$ becomes a Hilbert space with respect to the inner product
\begin{equation}\label{eq:51}
(f,g) \mapsto \int_X h( f, g) \mu = \int_X f(p)^\dagger \frac{\utilde{p}}{m} g(p) \hspace{0.1cm} d\mu (p).
\end{equation}

Substituting $V'$ in place of $V$ in the formula Eq.~(\ref{eq:27}), we obtain a unitary representation of the group $\mathbb{R}^4 \ltimes SL(2, \mathbb{C})$ on the Hilbert space $\mathcal{H}'$,

\begin{equation}\label{eq:52}
\left[U'(\Lambda, a ) \phi \right] (p) = e^{-i p_\mu a^\mu} \Lambda \phi (\Lambda^{-1} p).
\end{equation}

We see that this alternative representation space is free from the choice of $L(p)$. But, there is one drawback: $\mathcal{H}'$ is not a tensor product system due to the presence of the nontrivial inner product Eq.~(\ref{eq:51}). The $L(p)$-dependence has been subsumed into the nontrivial inner product. Conversely, to make the representation space into a tensor product system as in Eq.~(\ref{eq:1}), one must make a choice of $L(p)$.

This shows one advantage of the bundle description. Even though the representation space $\mathcal{H}'$ is not a tensor product system, we can talk about the discrete degree of freedom $\mathbb{C}^2$ freely since both $\mathcal{H}$ and $\mathcal{H}'$ have the same bundle $X \times \mathbb{C}^2$ as their underlying bundles from which they can be constructed.

Define a linear map $\alpha : \mathcal{H}' \rightarrow \mathcal{H}$ as
\begin{equation}\label{eq:53}
\left[ \alpha (f) \right] (p) = L^{-1} \circ f (p) = \sqrt{\frac{\utilde{p}}{m}} f(p)
\end{equation}
which is a Hilbert space isomorphism since $L^{-1}$ is an Hermitian bundle isomorphism. Using this, we can easily show that
\begin{equation}\label{eq:54}
U' ( \Lambda, a ) = \alpha \circ U(\Lambda , a) \circ \alpha^{-1},
\end{equation}
which implies that the just-defined representation $(U', \mathcal{H}')$ is unitarily equivalent to $(U, \mathcal{H})$.

We see that states $\psi \in \mathcal{H}$ and $\phi \in  \mathcal{H}'$ are fields of qubits distributed among all possible values of $p \in X$. The interpretations given in Sect.~\ref{sec:3F} suggest that while $\psi(p)$ has spin information which becomes available only when one is also given $L(p)$, the information contained in $\phi(p)$ is directly available from the fixed inertial observer who had prepared $\phi$ in the first place. This fact can also be seen from the following. If $\psi = \alpha (\phi)$, then
\begin{equation}\label{eq:55}
\psi (p) = L(p)^{-1} \phi (p) = \left[ U' (L(p)^{-1} ) \phi \right] (k)
\end{equation}
which implies that $\psi(p)$ is the spin information of $\phi$ as seen from an $L(p)^{-1}$-transformed observer.

\subsection{\label{sec:4A}The spin operator}

Let's examine the representation space $\mathcal{H}_0$ constructed from the bundle $E_0$. The observable
\begin{equation}\label{eq:56}
\mathbf{S} = \frac{1}{2} \boldsymbol{\tau}
\end{equation}
on the standard Hilbert space $\mathcal{H}_0 = L^2 (X , \mu) \otimes \mathbb{C}^2$ is the spin operator used in nonrelativistic Quantum Mechanics (e.g. Stern-Gerlach experiment). But, as explained in the introduction, this operator does not capture the momentum-independent spin information due to the inherent momentum dependence of the $\mathbb{C}^2$ component. Indeed, on the alternative representation space $\mathcal{H}'$, it becomes
\begin{equation}\label{eq:57}
\mathbf{S}' =  \alpha^{-1} \circ \mathbf{S} \circ \alpha.
\end{equation}

So, the momentum dependence of the operator Eq.~(\ref{eq:56}), which was hidden on the Hilbert space $\mathcal{H}_0$, becomes visible from the space $\mathcal{H}'$.

In the appendix, it is shown that $\mathbf{S}'$ is exactly the Newton-Wigner spin operator on the alternative space $\mathcal{H}'$, i.e.,
\begin{equation}\label{eq:58}
\mathbf{S}' = \frac{1}{m} \left( \mathbf{W} - \frac{W^0 \mathbf{P} }{ m + P^0 } \right)
\end{equation}
where $\mathbf{W}$ is the Pauli-Lubanski vector operator defined as the spatial component of the four-vector operator
\begin{equation}\label{eq:59}
W^\mu = \frac{1}{2} \varepsilon^{\nu \alpha \beta \mu} P_\nu J_{\alpha \beta}.
\end{equation}

This shows that the interpretation given in Sect.~\ref{sec:3F3} is also valid on the level of Hilbert space and operators.

Also, it is interesting to note that
\begin{equation}\label{eq:60}
W  = \frac{1}{2} \tilde{P} \tau - \frac{1}{2} P I
\end{equation}
on $\mathcal{H}'$ by Eqs.~(\ref{eq:A.2}) and (\ref{eq:A.11}).

Hence, for a state $\phi \in \mathcal{H}'$, the expectation value for the Pauli-Lubansky four-vector is given by
\begin{align}\label{eq:61}
\langle W^\mu \rangle_\phi = \int_X d\mu(p)  \left(\phi(p)^\dagger \frac{\utilde{p}}{m} \right) \left( \frac{1}{2} (\tilde{p} \tau^\mu - p^\mu I ) \right) \phi(p)  \nonumber \\
= m \int_X d\mu (p) \left(\phi(p)^\dagger (\frac{1}{2}  \tau^\mu ) \phi(p) \right) - \frac{1}{2} \langle P^\mu \rangle_\phi,
\end{align}
whereas the expectation value for the Newton-Wigner spin is
\begin{equation}\label{eq:62}
\langle \mathbf{S}_{NW} \rangle_\phi = \int_X d \mu(p) [\alpha \phi](p)^\dagger \left(\frac{1}{2} \boldsymbol{\tau}\right) [\alpha \phi ] (p).
\end{equation}
which are wave function analogues of Eqs.~(\ref{eq:46}) and (\ref{eq:49}), respectively.

Before we finish this subsection, we should mention that there is no universally accepted definition of relativistic spin operator. The Newton-Wigner spin operator, Eq.~(\ref{eq:58}), is just one of many proposed operators. Several promising candidates as well as their properties are thoroughly discussed in \cite{bauke2014, bauke2014b}. (Notice that our Newton-Wigner spin operator is referred to as the \textit{Pryce spin operator} there.) Interested readers are invited to test these various spin operators using the vector bundle formalism developed in this paper.

The lack of unanimity in the definition of spin operator stems from the fact that there is no universally agrred upon definition of relativistic position operator. This issue is addressed in \cite{terno2016}, which also suggests a measurement-based formalism to tackle the problem.

Finally, since this paper's primary concern is about inertial observers' perception of internal quantum states of a free particle, operational aspects of spin, such as its interaction with electromagnetic field, are not dicussed here. A manifestly Lorentz covariant treatment of this aspect of relativistic spin is discussed in \cite{DERIGLAZOV2016}.

\subsection{\label{sec:4B}The spin reduced density matrix}

Now, let's investigate the spin reduced density matrix of \cite{peres2002}. Let $\psi = f \chi \in \mathcal{H}$ where $f$ is a Schwarz class function (it was so chosen to avoid any integrability issue) such that $\int_X d \mu (p) |f(p)|^2 = 1$ and $\chi :X \rightarrow E$ is a measurable section such that $\|\chi(p)\|_{E_p}= 1$ for all $p \in X$, i.e. $\chi(p)$ is a qubit in $E_p$ for each $p \in X$. Recalling Eq.~(\ref{eq:43}), we see
\begin{equation}\label{eq:63}
\psi(p) \psi (p)^\dagger = \frac{|f(p)|^2}{2} ( \boldsymbol{\tau} \cdot \mathbf{n} (p) + I_2 )
\end{equation}
where $\mathbf{n}(p) = \chi (p)^\dagger \boldsymbol{\tau} \chi(p)$ is the spin direction of the qubit $\chi(p) \in E_p$. So, the spin reduced density matrix is
\begin{equation}\label{eq:64}
\int_X  \psi \psi^\dagger \mu = \frac{1}{2} + \int_X d \mu (p) |f(p)|^2 \frac{1}{2} \boldsymbol{\tau} \cdot \mathbf{n} (p) 
\end{equation}
which is just a weighted average of the spin $\frac{1}{2}\mathbf{n}(p)$ of the qubits $\chi(p)$. However, since each three-vector $\mathbf{n}(p)$ gets its meaning only with respect to the $L(p)^{-1}$-transformed frame (cf. Sect.~\ref{sec:3F1}), Eq.~(\ref{eq:64}) is a summation of vectors living in a whole lot of different coordinate systems. So, we see that this average value really has no meaning at all.

\cite{peres2002} asserted that this matrix is meaningless only because it does not transform covariantly under Lorentz transformations and \cite{saldanha2012a} gave a physical argument supporting the assertion. Our analysis has just given a mathematical proof for these earlier assertions. However, this proof has a stronger implication: It shows that the spin reduced density matrix of \cite{peres2002} is void of any meaning at all, not just in terms of Lorentz non-covariance or the impossibility of measurement.

Any anticipation that this would give the spin information independent of the momentum variable is an illusion caused by the form of the standard representation space Eq.~(\ref{eq:1}) as a tensor product system. Also, any attempt to gain whatever information from this partial trace operation seems to be meaningless either. For one thing, to use a tensor product system as the representation space, one has to make a certain choice for $L(p)$ (cf. Sect.~\ref{sec:4}), depending on which the $\mathbb{C}^2$ component would take a variety of meanings \cite{avelar2013}. And even if we have made a great choice (such as $L(p) = L_0 (p)$), the partial trace operation is bound to face the issue of summing over vectors living in different coordinate systems as can be seen in Eq.~(\ref{eq:64}). So, we totally discard the strategy to obtain spin information by means of the partial trace operation.

\subsection{\label{sec:4C}The Pauli-Lubansky reduced matrix}
Let's just accept Eq.~(\ref{eq:64}) as a kind of average operation and see if we can assign a meaning to it. First, notice that we can rewrite the quantity as
\begin{equation}\label{eq:65}
 \int_X d \mu (p)  \frac{|f(p)|^2}{2}\left((\mathbf{n}(p))^\sim + \frac{ \tilde{k}}{m} \right).
\end{equation}

Since summing over vectors living in different coordinate systems is meaningless, we can give this average operation a substance only by first pulling the integrand vectors back to a fixed coordinate system by applying $L(p)$ and then executing the summation. The result is
\begin{align}\label{eq:66}
\int_X d \mu (p) \frac{|f(p)|^2 }{2} \left( (L(p) \mathbf{n}(p) )^\sim + \frac{ (L(p) k)^\sim }{m} \right) \nonumber \\
 = \int_X d\mu (p) |f(p)|^2  \left( \frac{\tilde{w}(p)}{m} + \frac{\tilde{p}}{2m} \right).
\end{align}

We notice that this is just the following operation defined on $\mathcal{H}'$
\begin{equation}\label{eq:67}
\sigma (\phi) := \int_X \phi(p) \phi(p)^\dagger d\mu(p), \footnote{This operation, however, is defined only on $\phi \in \mathcal{H}'$ for which the integral exists, which is the case, for example when the component functions of $\phi$ are Schwarz class functions.}
\end{equation}
applied to $\alpha^{-1} (\psi) \in \mathcal{H}'$ (cf. Eq.~(\ref{eq:48})). By Eq.~(\ref{eq:46}), the modified matrix has information about the average of the four-vector $w + \frac{p}{2}$. We will call this matrix the \textit{Pauli-Lubansky reduced matrix} in this paper.

It is very important to notice that Eq.~(\ref{eq:67}) is not a partial trace operation since, for one thing, $\mathcal{H}'$ is not a tensor product system and for another, $\phi(p) \phi(p)^\dagger $ is not the usual density matrix corresponding to the vector $\phi (p) \in E'_p$, due to the form of the inner product Eq.~(\ref{eq:29}) on $E'_p$.

Here are some features of this matrix. $\sigma = \sigma (\phi)$ is positive and has a nonzero trace:
\begin{equation}\label{eq:68}
u^\dagger \sigma u = \int_X d \mu(p) \hspace{0.1cm} u^\dagger \phi(p) \phi(p)^\dagger u \geq 0 \quad \forall u \in \mathbb{C}^2
\end{equation}
and
\begin{equation}\label{eq:69}
\text{Tr} \hspace{0.1cm} \sigma = \int_X d \mu(p) \left( |\phi_1 (p)|^2 + |\phi_2 (p)|^2 \right) > 0.
\end{equation}

So, the matrix $\sigma$ can be normalized to yield a density matrix. Let's find its transformation law under a change of reference frame.

Suppose Alice has prepared a state $\phi \in \mathcal{H}'$ and formed the matrix $\sigma_A = \sigma (\phi)$. Consider another observer Bob, in whose frame the state is $U' (\Lambda, a ) \phi \in \mathcal{H}'$. Then, due to the transformation law Eq.~(\ref{eq:52}) and the Lorentz invariance of $\mu$ (Eq.~(\ref{eq:10})), we have
\begin{equation}\label{eq:70}
\sigma_B = \int_X d\mu (p) \hspace{0.1cm} \Lambda \phi ( \Lambda^{-1} p) \phi( \Lambda^{-1} p)^\dagger \Lambda^{\dagger} = \Lambda \sigma_A \Lambda^{\dagger}.
\end{equation}

So, we see that the Pauli-Lubansky reduced matrix is Lorentz covariant and hence has a relativistically invariant meaning. When the average momentum $\langle p \rangle_\phi$ of the state $\phi \in \mathcal{H}'$ is available, the matrix
\begin{equation}\label{eq:71}
\theta:= m\sigma - \frac{1}{2} (\langle p \rangle_\phi )^\sim
\end{equation}
gives information about the average Pauli-Lubansky four-vector of the particle represented by the state $\phi$ (cf. Eq.~(\ref{eq:46})). This is also Lorentz covariant in the sense of Eq.~(\ref{eq:70}) since $\sigma$ and $(\langle p \rangle_\phi )^\sim$ follow the same transformation law.

It seems that this is as far as we can go in an endeavor to give the spin reduced density matrix Eq.~(\ref{eq:64}) a meaning. The matrix Eq.~(\ref{eq:67}) first appeared in \cite{caban2005} with the name "Lorentz covariant reduced spin density matrix". We have just shown that this is the only way that the spin reduced density matrix Eq,~(\ref{eq:64}) can be modified to get any significance.

Operational aspects of the Pauli-Lubansky reduced matrix is investigated in \cite{caban2005} and also briefly discussed in \cite{terno2016}. \footnote{Although our representation space is different from the Dirac bispinor representation used in \cite{caban2005}, our Pauli-Lubansky reduced matrix Eq.~(\ref{eq:66}) and the reduced matrix of \cite{caban2005} contain the same information. See (32) of \cite{caban2018} where the matrix is written explicitly.}

\section{\label{sec:5}Remarks}

Here are some remarks on the subjects dealt with in this paper. Although the vector bundle point of view is new, representations similar to the alternative representation (Sect.~\ref{sec:4}) have already been studied (either implicitly or explicitly) in the literature.

\cite{caban2005} used it to construct the Pauli-Lubansky reduced matrix Eq.~(\ref{eq:67}) (which was called the Lorentz covariant reduced spin density matrix in that paper). \cite{caban2012} used this representation space by the name "covariant basis" and proved Eq.~(\ref{eq:58}). See also \cite{terno2016} for a treatment of the same matrix.

Unlike the present paper, however, the two papers used the Dirac bispinor representation to describe a massive spin-1/2 particle. In fact, in this description, the equivalence Eq.~(\ref{eq:53}) between the standard and the alternative representations becomes the famous \textit{Foldy–Wouthuysen transformation} (cf. \cite{ deriglazov2020}),\footnote{I appreciate Alexei Deriglazov for pointing this out to me.} which has been extensively studied in the literature.

Very recently, the alternative representation space was taken up again for the study of a Lorentz covariant, unitary time evolution of a uniformly accelerating particle (\cite{caban2018, caban2019}). The relativistic chiral qubits considered in these two papers can be identified as elements of the $E'$-bundle of this paper.

In all these papers, the simplicity of the transformation law Eq.~(\ref{eq:52}) seems to be the main motivation for its introduction. In addition to this good feature, we showed, in this paper, that this representation is more fundamental than all the other representations in that it is free from the choice of the boostings $L(p)$, which caused the ambiguity in the meaning of the spin of a moving particle and the Lorentz-noncovariance of the spin reduced density matrix Eq.~(\ref{eq:64}), and that the values of the wave functions of this representation space are what fixed inertial observers "perceive as apparent information" in their rest frames (cf. Sect.~\ref{sec:3F}).

\section{\label{sec:6}Conclusions}
A vector bundle theoretic way to view the single-particle state space of a massive spin-1/2 particle was introduced in this paper. The vector bundles were thought of as an assembly of two-level systems corresponding to all possible values of $p$ in the mass shell. This viewpoint helped us to remove the $L(p)$-arbitrariness issue first recognized in \cite{avelar2013}. Two kinds of bundles were considered, the $E$-bundle and the $E'$-bundle. The $E$-bundle was related to the standard representation space Eq.~(\ref{eq:1}) and hence dependent on the choice of $L(p)$. But, the $E'$-bundle was completely free from this.

In this viewpoint, every inertial observer is given a bundle (either $E$ or $E'$) for the description of a massive spin-1/2 particle. The transformation laws for the bundles have been explicitly obtained and a physical interpretation for each bundle has been given. In this interpretation, the qubits in the fibers of the $E'$-bundle are directly accessible from the rest frame and give information about the Pauli-Lubansky vector of the particle. But, those of the $E$-bundle are not accessible from the rest frame unless the observer is also given the knowledge of $L(p)$. But, once we make the standard choice $L(p)=L_0 (p)$ (cf. \cite{weinberg}), then the qubits in the fibers of the resulting bundle $E_0$ have information about the Newton-Wigner spin.

There are Hilbert space descriptions that naturally come out from this bundle description, namely, the $L^2$-section descriptions of the corresponding bundles. The one associated with the $E$-bundle is just the standard description of the single-particle state space used in the RQI literature. The other one associated with the $E'$-bundle is an alternative representation space, which is also free from the $L(p)$-arbitrariness issue. The physical interpretations for the bundles are still valid on the level of Hilbert space and operators.

Using the bundle point of view, we also gave a mathematical proof as to why the spin reduced density matrix introduced in \cite{peres2002} is meaningless, as claimed in \cite{saldanha2012a}. In fact, we showed that any attempt to obtain information about the spin of the particle by means of the partial trace operation is bound to fail because it cannot avoid the issue of summing over vectors which reside in different coordinate systems. The only way to make this matrix meaningful is to pull back the integrands to a fixed coordinate system so that they become summable. The result is the Pauli-Lubansky reduced matrix, which contains information about the average Pauli-Lubansky four-vector plus the four-momentum of the particle.

\begin{acknowledgements}
H. Lee was supported by the Basic Science Research Program through the National Research Foundation of Korea (NRF) Grant NRF-2017R1E1A1A03070510.
\end{acknowledgements}

\section*{Data Availability Statement}

Data sharing is not applicable to this article as no new data were created or analyzed in this study.

\appendix*
\section{}

Denoting \textit{purely symbolically}
\begin{equation}\label{eq:A.1}
	x^\mu := \frac{1}{i} \frac{\partial}{\partial p_\mu },
\end{equation}
the following are the 10 standard mechanical operators :
\begin{subequations}
\label{eq:A.2}
\begin{eqnarray}
	\left[P^0 \psi \right] (p) = \left[ i \frac{\partial}{\partial t} U' ( I, t e_0 )  \psi \right] (p)  = p^0 \psi (p) \hspace{0.5cm}  \\
	\left[P^j \psi \right] (p) =   \left[ - i \frac{\partial}{\partial x } U'(I, x e_j) \psi \right] (p) =  p^j \psi (p) \hspace{0.5cm}
\end{eqnarray}
\end{subequations}
\begin{eqnarray}\label{eq:A.3}
	\left[ J^j \psi \right] (p) &=&  \left[ i \frac{\partial}{\partial \theta} U' \left(R^j (\theta) , 0 \right) \psi \right] (p) \hspace{1.2cm} \nonumber \\
 &=& \left[ \frac{1}{2} \tau^j  - \left( \mathbf{p} \times \mathbf{x} \right)^j \right] \psi (p)
\end{eqnarray}
\begin{eqnarray}\label{eq:A.4}
\left[K^j \psi \right] (p) &=& \left[ i \frac{\partial}{\partial u} U' \left( B^j (u) , 0 \right) \psi \right] (p) \nonumber \\ 
 &&= \left[ \frac{i}{2} \tau^j  - \left( p^0 x^j - p^j x^0 \right) \right] \psi (p)
\end{eqnarray}
where the standard rotation and boosting matrices $R^j (\theta)$ and $B^j (u)$ are given by
\begin{subequations}\label{eq:A.5}
\begin{eqnarray}
R^j (\theta) &=& \exp \left( - \frac{i}{2} \theta \tau^j  \right) \\
B^j (u) &=& \exp \left( \frac{1}{2} u \tau^j \right)
\end{eqnarray}
\end{subequations}
respectively.

Now, inserting these formulae into Eq.~(\ref{eq:59}) while recalling $J^i = \varepsilon_{ijk} J^{jk}$ and $K^j = J^{0j}$ (cf. \cite{ryder}), we obtain
\begin{subequations}\label{eq:A.6}
\begin{eqnarray}
W^0 = \mathbf{P} \cdot \mathbf{J} =  \frac{1}{2} \mathbf{p} \cdot \boldsymbol{\tau}  = - \frac{1}{2} ( \utilde{p} - p^0 I ) \\
\mathbf{W} = P^0 \mathbf{J} - \mathbf{P} \times \mathbf{K}  =  \frac{1}{2} \left( p^0 \boldsymbol{\tau}  - i \mathbf{p} \times \boldsymbol{\tau}  \right) 
\end{eqnarray}
\end{subequations}
which, after a bit of algebra, becomes
\begin{equation}\label{eq:A.7}
\mathbf{W} = \frac{1}{2} \left( \boldsymbol{\tau} \utilde{p} + \mathbf{p} I \right).
\end{equation}

Inserting Eqs.~(\ref{eq:A.2}), (\ref{eq:A.6}) and (\ref{eq:A.7}) into the RHS of Eq.~(\ref{eq:58}) yields
\begin{eqnarray}\label{eq:A.8}
\mathbf{S}_{NW} := \frac{1}{2m} \left( \boldsymbol{\tau} \utilde{p} + \mathbf{p}I  + \mathbf{p} \frac{\utilde{p} - p_0 I }{m + p_0}  \right) \nonumber \\
= \frac{1}{2m} \left( \boldsymbol{\tau} \utilde{p} + \mathbf{p} \frac{ \utilde{p} + m I }{ m + p_0} \right).
\end{eqnarray}

By the way, using Eq.~(\ref{eq:14}) and the commutation relation $\tilde{p} \boldsymbol{\tau} = \boldsymbol{\tau} \utilde{p} + 2 \mathbf{p} I$ of the Pauli matrices, Eq.~(\ref{eq:57}) becomes
\begin{eqnarray}\label{eq:A.9}
\mathbf{S}' &=& \sqrt{\frac{\tilde{p}}{m}} \mathbf{S} \sqrt{\frac{\utilde{p}}{m}} = \frac{1}{2 \sqrt{2m( m + p_0 ) }} \left( \tilde{p} + mI \right) \boldsymbol{\tau} \sqrt{\frac{\utilde{p}}{m}} \nonumber \\
&&= \frac{1}{2 \sqrt{2m (m+p_0 )} } \left( \boldsymbol{\tau} (\utilde{p} + mI)  + 2 \mathbf{p} I \right) \sqrt{\frac{\utilde{p}}{m}} \nonumber \\
&&= \frac{1}{2m} \left( \boldsymbol{\tau} \utilde{p} + \mathbf{p} \frac{\utilde{p} + mI}{m + p_0} \right)
\end{eqnarray}
which is exactly Eq.~(\ref{eq:A.8}). Therefore, we conclude
\begin{equation}\label{eq:A.10}
\mathbf{S}' = \mathbf{S}_{NW}.
\end{equation}

By the way, using the commutation relation $\boldsymbol{\tau}\utilde{p} = - 2 \mathbf{p} I + \tilde{p} \boldsymbol{\tau}$ (cf. Eq.~(\ref{eq:13})), we can summarize Eqs.~(\ref{eq:A.6}) and (\ref{eq:A.7}) by the following formula
\begin{equation}\label{eq:A.11}
W^\mu = \frac{1}{2} \left( \tilde{p} \tau^\mu - p^\mu I \right).
\end{equation}
\nocite{*}

\bibliography{aipsamp}

\end{document}